\documentclass[11pt]{article}

\usepackage{float}
\usepackage[a4paper,margin=1in]{geometry}
\usepackage{graphicx}
\usepackage{amsmath}
\usepackage{hyperref}
\usepackage{authblk}

\title{\textbf{Chhavi: A Python Tool for Converting RAMSES Outputs to VTKHDF for High-Fidelity AMR Visualization}}

\author[1,2]{Hemangi C. Varkal}
\author[1]{Shubhankar R. Gharote}
\author[1]{Munn Vinayak Shukla}
\author[1]{Mehul Pandya}

\affil[1]{Space Applications Centre (SAC), ISRO, Ahmedabad, India}
\affil[2]{L. J. Institute of Engineering and Technology, Ahmedabad, India}

\date{}

\begin{document}

\maketitle
\vspace{-1.0em}

\begin{abstract}
RAMSES is an adaptive mesh refinement (AMR) astrophysical simulation code that generates high-resolution multiscale data. However, its native binary output format is not directly compatible with standard visualization tools, making efficient analysis challenging. We present \textit{Chhavi}, an open-source Python tool that converts RAMSES outputs into the VTKHDF format for direct use in visualization platforms such as ParaView. The tool reconstructs the AMR hierarchy, preserves key physical fields including density, pressure, and velocity, and organizes them into a structured representation suitable for multi-resolution visualization. 

The conversion is validated using the three-dimensional Sedov blast wave test case. Quantitative evaluation through radial profile comparison and Lin’s concordance correlation coefficient demonstrates strong agreement between the original and converted datasets, confirming both physical and structural fidelity. Chhavi provides a scalable bridge between RAMSES simulation outputs and modern visualization workflows, supporting efficient and reproducible analysis in computational astrophysics.
\end{abstract}

\section{Introduction}

Astrophysical simulations are essential for studying the formation and evolution of structures in the universe, including interstellar filaments, star-forming regions, and galactic environments. Modern simulation frameworks such as RAMSES~\cite{ramses} employ adaptive mesh refinement (AMR) to capture multiscale physical processes by dynamically increasing resolution in regions of interest. This results in hierarchical datasets with high physical fidelity, but also introduces challenges for visualization and interpretation.

The hierarchical structure of AMR complicates the representation of spatial relationships across refinement levels, particularly in interactive analysis and visualization workflows. Effective visualization is therefore crucial for examining structural features, validating physical behavior, and exploring simulation outputs. However, RAMSES data are not directly compatible with standard visualization pipelines or widely supported formats such as VTK-based representations, often requiring custom conversion scripts or analysis-specific workflows.

Several tools, including \textit{yt}~\cite{yt} and \textit{Osyris}~\cite{osyris}, support the extraction and analysis of RAMSES data within scripting environments. While these tools are effective for in-memory analysis, they are not designed to provide a straightforward pathway for exporting data into formats suitable for integration with external visualization platforms such as ParaView~\cite{paraview}. This creates a gap between simulation outputs and visualization-ready data products.

To address this limitation, we introduce \textit{Chhavi}, a Python-based tool for converting RAMSES outputs into structured, visualization-ready representations while preserving the AMR hierarchy and essential physical fields. Chhavi enables seamless integration of RAMSES data with interactive visualization environments and supports reproducible computational astrophysics workflows.

The main contributions of this work are summarized as follows:
\begin{itemize}
    \item A modular pipeline for converting RAMSES outputs into visualization-compatible formats while preserving AMR hierarchy.
    \item Preservation of key physical fields required for scientific analysis.
    \item Validation of the conversion process using a standard astrophysical test case.
    \item A reproducible workflow for integrating RAMSES data with interactive visualization tools.
\end{itemize}

\section{Background and Related Work}

RAMSES~\cite{ramses} is a widely used astrophysical simulation code that employs adaptive mesh refinement (AMR) to model multiscale physical processes. Its data are organized as a hierarchy of grids with varying spatial resolution, where finer grids are embedded within coarser ones based on physical criteria. This multilevel structure is central to its efficiency but introduces complexity in downstream data handling and visualization.

Scientific visualization frameworks such as the Visualization Toolkit (VTK)~\cite{vtk} and ParaView~\cite{paraview} provide support for multiresolution datasets through data models such as \textit{Overlapping AMR}. More recently, the VTKHDF format has been introduced to represent such data within a compact HDF5-based structure, enabling efficient storage and scalable visualization. These frameworks require input data to conform to well-defined formats that encode both spatial organization and associated physical fields. The HDF5 library provides the underlying hierarchical storage model commonly used for such structured scientific data~\cite{hdf5}.

For RAMSES outputs, tools such as \textit{yt}~\cite{yt} and \textit{Osyris}~\cite{osyris} provide mechanisms for reading simulation data and accessing AMR structures and physical quantities. These tools are primarily designed for analysis within Python environments, offering functionality for slicing, projection, and statistical evaluation. While they enable flexible data exploration, exporting the data into formats compatible with external visualization frameworks typically requires additional processing or custom implementations.

As a result, there is no straightforward and standardized pathway for converting RAMSES outputs into formats that can be directly consumed by scientific visualization frameworks while preserving the AMR hierarchy and associated physical fields. This limitation motivates the need for a dedicated conversion pipeline that bridges simulation outputs with visualization-ready data representations.

\section{Methodology}

\begin{figure}[t]
    \centering
    \includegraphics[width=1.04\columnwidth]{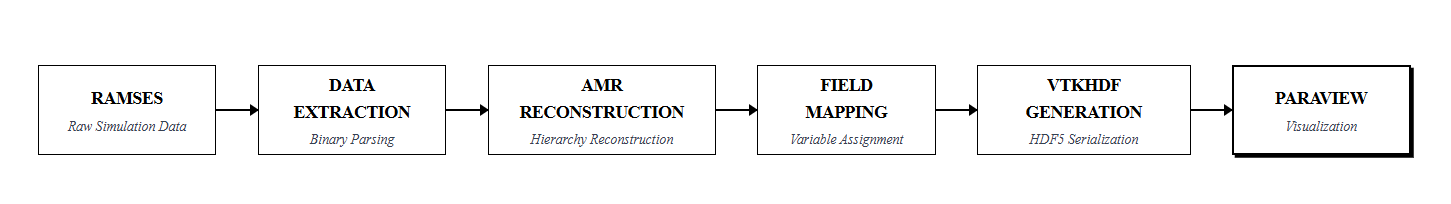}
    \caption{Overview of the Chhavi pipeline. RAMSES simulation outputs are processed through data extraction, AMR hierarchy reconstruction, and field mapping, and are then converted into the VTKHDF format for visualization in ParaView.}
    \label{fig:pipeline}
\end{figure}

The Chhavi pipeline converts RAMSES simulation outputs into a structured format suitable for visualization while preserving the adaptive mesh refinement (AMR) hierarchy and associated physical fields. The workflow consists of four main stages: data extraction, hierarchy reconstruction, field mapping, and output generation. An overview of the pipeline is shown in Fig.~\ref{fig:pipeline}.

\subsection{Data Extraction}

The conversion process begins by reading RAMSES output files and extracting the simulation data required for visualization, including cell-centered spatial coordinates, refinement levels, and physical quantities such as density, pressure, and velocity. This stage collects the data in a consistent representation that can be processed independently of the original file structure.

\subsection{AMR Hierarchy Reconstruction}

RAMSES data are inherently hierarchical, with cells distributed across multiple refinement levels. In this stage, the extracted data are organized according to their refinement levels and spatial relationships to reconstruct the multilevel grid structure. This step preserves the nested organization of the AMR grid and ensures that the spatial hierarchy is retained in the converted representation.

\subsection{Field Mapping}

After reconstructing the hierarchical structure, scalar and vector fields are associated with the corresponding cells in the grid. Each physical quantity is mapped directly to its respective cell without interpolation or transformation, ensuring that the original simulation values are preserved and remain consistent across refinement levels.

\subsection{VTKHDF Generation}

In the final stage, the processed data are written to the VTKHDF format. The reconstructed hierarchy and associated fields are encoded within an HDF5-based structure using hierarchical grouping that reflects the AMR levels. This representation is designed for compatibility with modern visualization frameworks and supports efficient access and multi-resolution visualization.

The modular design of the pipeline allows each stage to operate independently, improving extensibility and enabling adaptation to different simulation configurations and data requirements.

\section{Implementation}

Chhavi is implemented as a Python-based command-line tool that provides an efficient and reproducible pipeline for converting RAMSES simulation outputs into VTKHDF format. The implementation follows the modular workflow shown in Fig.~\ref{fig:pipeline}, with each stage corresponding to a distinct processing component.

\subsection{System Design}

The pipeline is structured as a set of independent modules corresponding to the main stages of the conversion process. RAMSES output directories are parsed to identify available simulation snapshots, and the relevant files are processed based on user input. This modular design allows individual components to be extended or modified without affecting the overall workflow.

\subsection{Command-Line Interface}

Chhavi is executed through a command-line interface that enables users to specify input and output configurations. Typical parameters include the base directory of the RAMSES outputs, the simulation folder containing the desired output, the target snapshot number, the output file prefix, the set of physical fields to be extracted, the output directory, and the number of worker processes for parallel conversion.

A typical usage example is shown below:

\begin{verbatim}
chhavi --base-dir ramses_outputs/ \
       --folder-name sedov_3d/ -n 1 \
       --output-prefix sedov_test \
       --fields density,velocity,pressure \
       --dry-run \
       --output-dir ./vtk_outputs \
       --nproc 1
\end{verbatim}

This command selects a RAMSES output from the specified simulation folder, extracts the chosen physical fields, and generates a corresponding VTKHDF file in the target directory. The implementation also supports execution through a Python module interface for programmatic usage.

\subsection{Execution Workflow}

The pipeline operates sequentially across its processing stages, with intermediate data maintained in memory to minimize disk I/O. Data are organized according to refinement levels and passed between modules in a structured form, ensuring consistency throughout the conversion process. This design enables efficient handling of large AMR datasets while preserving the integrity of the hierarchical structure.

\subsection{Performance Considerations}

The implementation is designed to handle large simulation datasets efficiently by minimizing redundant computations and maintaining streamlined data flow between pipeline stages. Parallel conversion is supported through the \texttt{--nproc} option, allowing users to configure the number of worker processes based on available computational resources.

Overall, Chhavi provides a practical and extensible framework for converting RAMSES outputs into VTKHDF representations, supporting scalable and reproducible scientific workflows.

\subsection{Availability}

The source code and detailed implementation of Chhavi are publicly available at \url{https://github.com/HemangiVarkal/Chhavi}. The tool is also distributed via PyPI at \url{https://pypi.org/project/chhavi/}, enabling straightforward installation and usage.

\section{Results and Validation}

The proposed conversion pipeline is evaluated using the three-dimensional Sedov blast wave test case, a standard benchmark for validating hydrodynamic simulations. This test case consists of a spherically symmetric shock wave propagating outward from an initial point explosion, making it well-suited for assessing both structural and physical fidelity in the converted data.

\subsection{Qualitative Assessment}

The converted VTKHDF outputs were visualized using ParaView~\cite{paraview} to assess the preservation of spatial structure and physical features. The resulting visualizations reproduce the expected spherical symmetry of the blast wave and clearly resolve the shock front across multiple refinement levels. The AMR hierarchy is retained, enabling consistent multiresolution rendering without visible discontinuities between refinement levels.

These results demonstrate that the conversion pipeline maintains the geometric and structural integrity of the original RAMSES data and supports effective visualization of complex simulation outputs.

\begin{figure}[H]
    \centering
    \includegraphics[width=\columnwidth]{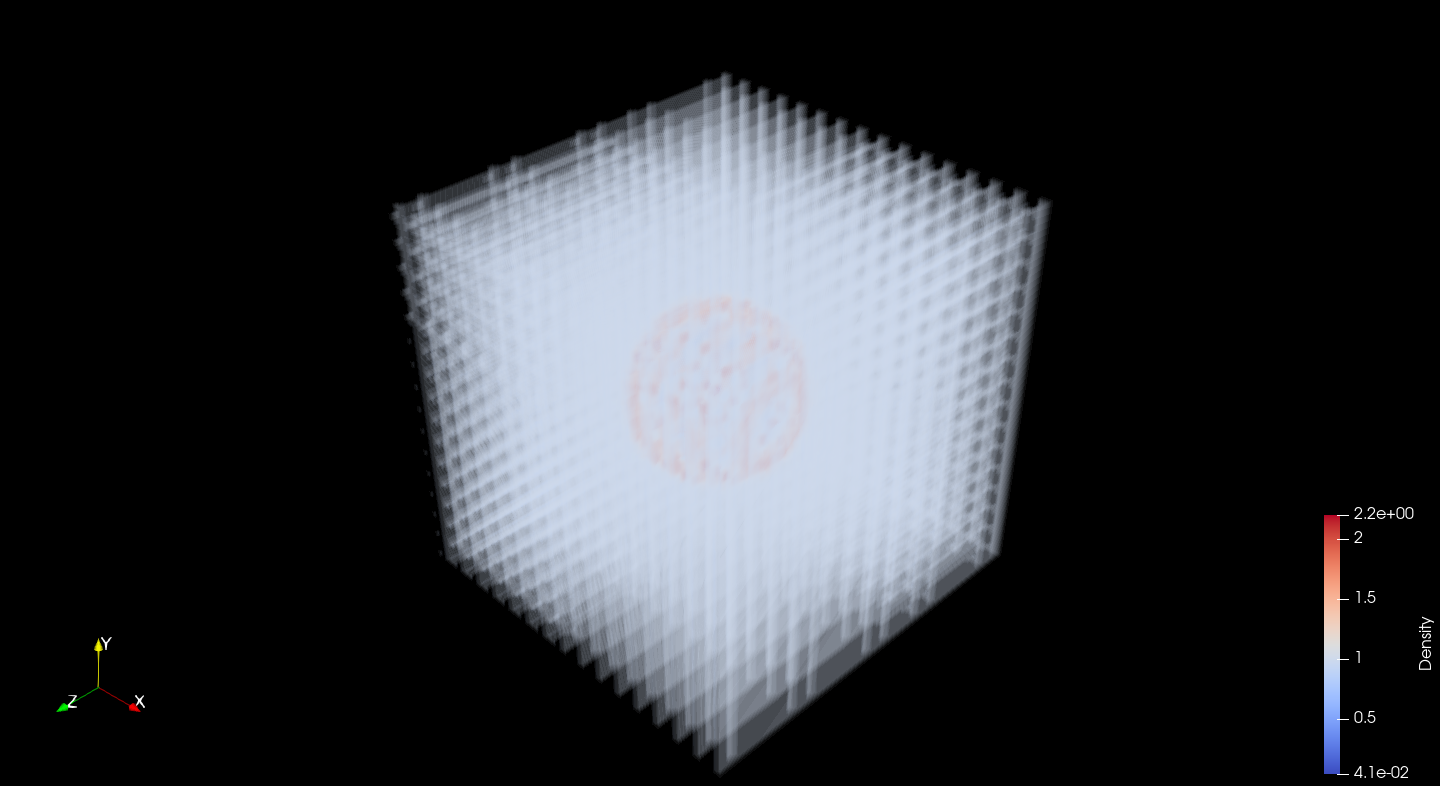}
    \caption{Visualization of the Sedov blast wave using the converted VTKHDF data. The spherical shock front is clearly resolved while preserving the multilevel AMR structure.}
    \label{fig:sedov}
\end{figure}

\subsection{Quantitative Validation}

To evaluate physical fidelity, radial profiles of key physical quantities were computed and compared between the original RAMSES outputs and the corresponding VTKHDF representations. The comparison focuses on variables such as density and pressure, which are essential for characterizing the shock structure.

Agreement between the original and converted datasets was quantified using Lin’s concordance correlation coefficient (CCC)~\cite{lin_ccc}. The computed CCC value was approximately 1.0, indicating near-perfect agreement between the original and converted radial profiles. This result suggests that the conversion process preserves both the structural and physical characteristics of the simulation data with negligible loss of information.

\begin{figure}[H]
    \centering
    \includegraphics[width=\columnwidth]{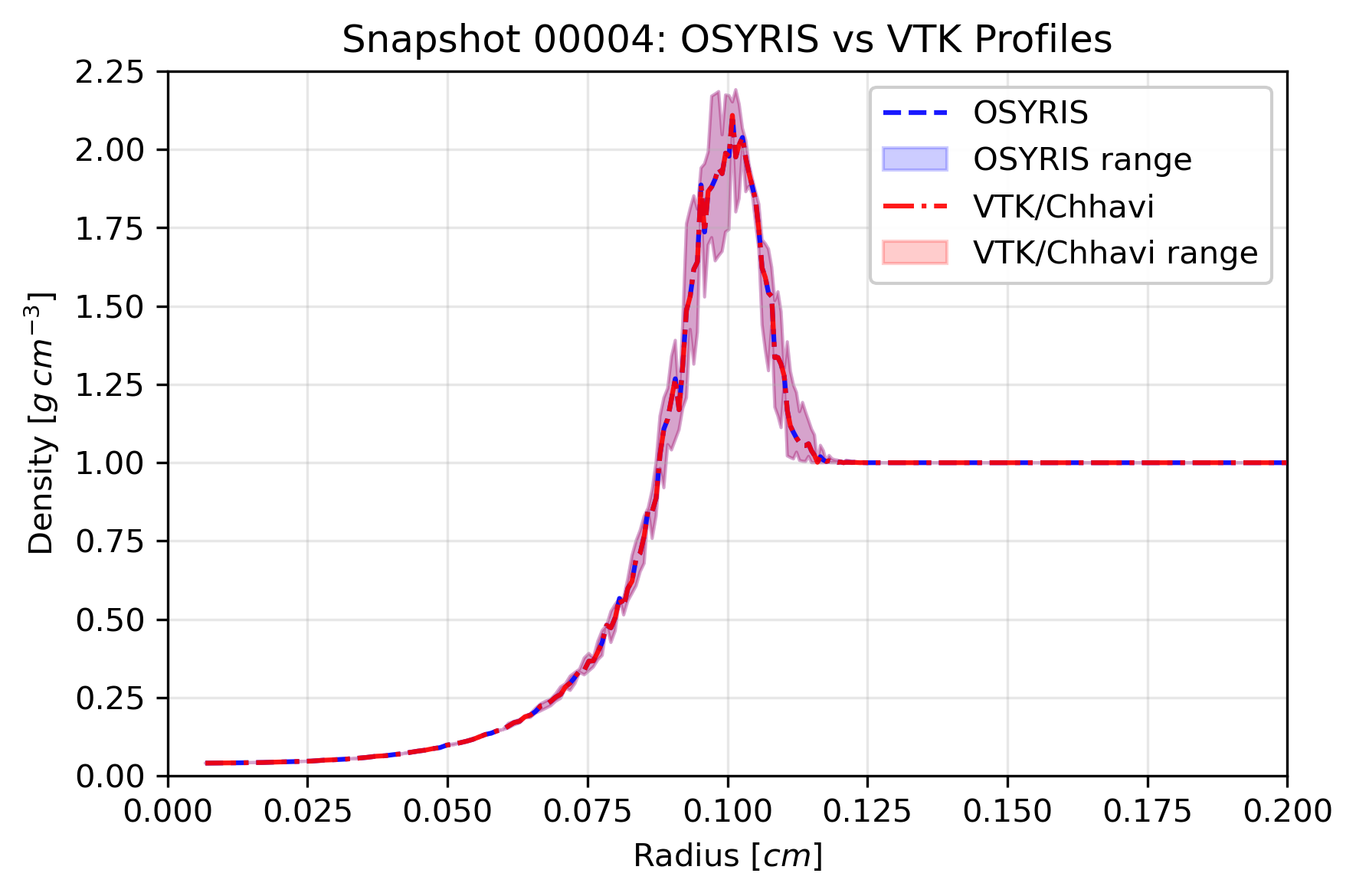}
    \caption{Radial profile comparison between the original RAMSES output and the converted VTKHDF representation for the Sedov blast wave test case. The close overlap of the profiles indicates preservation of the physical structure during conversion.}
    \label{fig:radial}
\end{figure}

\section{Discussion}

The results indicate that Chhavi provides a practical and reliable pathway for converting RAMSES simulation outputs into visualization-ready formats while preserving both structural and physical fidelity. Retaining the AMR hierarchy without interpolation ensures that multiscale features are accurately represented in the converted data.

A key advantage of the proposed approach lies in its compatibility with existing scientific visualization frameworks. By generating VTKHDF outputs, Chhavi enables direct integration with tools such as ParaView, allowing users to explore complex AMR datasets interactively without requiring custom post-processing pipelines. This improves accessibility and supports more efficient analysis workflows in computational astrophysics.

The modular design of the pipeline further enhances its flexibility. Individual components can be extended to support additional physical fields or adapted to different simulation configurations, making the tool applicable to a broader range of use cases beyond the specific test case considered in this work.

However, the current implementation has certain limitations. Validation is performed on a single benchmark test case, and additional evaluation on more complex simulations would strengthen the generality of the approach. While the conversion process itself is efficient, visualization of the resulting high-resolution AMR datasets can be computationally demanding. In particular, large datasets may require substantial memory and GPU resources for interactive exploration in visualization environments such as ParaView. This reflects the inherent complexity of multiresolution data rather than a limitation of the conversion process itself. Although the pipeline supports parallel execution, its performance may depend on system configuration and dataset size, motivating more detailed benchmarking in future work.

Overall, these findings demonstrate that Chhavi effectively bridges the gap between RAMSES simulation outputs and modern visualization workflows, providing a practical solution for high-fidelity AMR data conversion.

\section{Conclusion}

We presented Chhavi, a Python-based tool for converting RAMSES simulation outputs into the VTKHDF format for visualization. The proposed pipeline preserves the adaptive mesh refinement (AMR) hierarchy and associated physical fields, enabling accurate representation of multiscale simulation data in interactive visualization environments.

Validation on the three-dimensional Sedov blast wave test case demonstrated that the converted data retain both structural and physical fidelity, as confirmed by qualitative visualization and quantitative comparison. These results indicate that the conversion pipeline is reliable and suitable for scientific applications.

Chhavi bridges the gap between RAMSES data and modern visualization frameworks, providing a scalable and extensible solution for high-fidelity AMR data conversion and supporting reproducible research in computational astrophysics.

\section{Future Work}

Future work will focus on extending the validation of Chhavi to more complex and large-scale astrophysical simulations, including datasets with deeper AMR hierarchies and a wider range of physical conditions. This will further establish the generality and robustness of the conversion pipeline.

In addition, detailed performance benchmarking will evaluate scalability across different system configurations and dataset sizes. Optimizations for memory usage and parallel processing will also be explored to improve efficiency when handling large datasets.

Another direction involves improving integration with visualization workflows by developing more efficient strategies for handling high-resolution data, particularly in resource-constrained environments. Support for additional physical fields and enhanced customization options may further increase the flexibility of the tool.

Overall, these efforts aim to strengthen the applicability of Chhavi for large-scale scientific analysis and broaden its usability within the computational astrophysics community.

\bibliographystyle{plain}

\end{document}